\documentclass[twocolumn]{optica-article}

\usepackage[normalem]{ulem}
\usepackage{braket}
\usepackage{threeparttable}
\usepackage{graphicx}
\usepackage{multirow}
\usepackage{booktabs}
\usepackage{siunitx}
\usepackage{caption}
\usepackage{xcolor}

\newcommand{\add}[1]{#1}
\newcommand{\del}[1]{}
\newcommand{\todo}[1]{}

\journal{opticajournal}
\articletype{Research Article}

\begin{document}

\title{Inter-city 18.9 km free-space QKD over a dense urban corridor}

\author{%
{Ignacio H. López Grande}\textsuperscript{1,$\dagger$\add{,$\ddagger$}},
{Nicolas Linale}\textsuperscript{1,$\dagger$\add{,$\S$}},
{Giacomo Paganini}\textsuperscript{1},
{Achraf El Bouazzaoui}\textsuperscript{1},
{Adrián S. Blanca}\textsuperscript{1},
{Ilektra Karakosta-Amarantidou}\textsuperscript{1},
{Manu Canals i Codina}\textsuperscript{1\add{,$\P$}},
{Adam Vallés}\textsuperscript{2},
{Andrea Gorni}\textsuperscript{1},
{Jorge Fuenzalida}\textsuperscript{1},
{Evelyn A. Ortega}\textsuperscript{1}, and
{Valerio Pruneri}\textsuperscript{1,3}
}
\affil{\textsuperscript{1}ICFO-Institut de Ciències Fotòniques, The Barcelona Institute of Science and Technology, Castelldefels,
Barcelona, 08860, Spain}

\affil{\textsuperscript{2}Departament de Física, Universitat Autònoma de Barcelona, E-08193 Bellaterra, Spain}
\affil{\textsuperscript{3}ICREA-Institucio Catalana de Recerca i Estudis Avançats, 08010 Barcelona, Spain}
\affil{\textsuperscript{$\dagger$}These authors contributed equally to this work}
\affil{\add{\textsuperscript{$\ddagger$}Current affiliation: Luxquanta Technologies S.L., Av. Joan Carles I, 30, 1º1ª, 08908 L'Hospitalet de Llobregat, Barcelona, Spain}}
\affil{\add{\textsuperscript{$\S$}Current affiliation: Hispasat SA, Avda. de Bruselas, 35, 28108 Alcobendas, Madrid, Spain}}
\affil{\add{\textsuperscript{$\P$}Current affiliation: Lace Lithography AS, Espehaugen 33, 5258 Blomsterdalen, Bergen, Norway}}

\email{ignacio.lopez@alumni.icfo.eu}

\begin{abstract}
Terrestrial free-space quantum key distribution (QKD) provides a flexible complement to fiber-based networks, but its deployment over long horizontal paths in dense metropolitan environments remains challenging, as atmospheric turbulence, demanding pointing and tracking requirements, and operational constraints can substantially increase system complexity. We experimentally demonstrate the operation of a decoy-state BB84 system over an 18.9~km free-space link across a densely populated urban area in the metropolitan region of Barcelona  using an aviation-compliant optical architecture. The system operates at 785~nm using room-temperature single-photon detectors and requires no adaptive optics at the receiver. The modular QKD transmitter, based on a deployment-oriented design, is interfaced with a versatile free-space optical terminal and an independent GPS-disciplined synchronization architecture, facilitating its integration into heterogeneous optical infrastructures. 
We present the transmitter performance and optical-link characteristics, together with a 40-minute field test characterized by an average channel attenuation of approximately \del{18}\add{19}~dB and a mean QBER of 1.31\%. From the measured channel parameters, we estimate an average finite-size secret key rate exceeding 11~kbit/s. These results demonstrate the feasibility of inter-city free-space QKD over dense urban areas using a comparatively simple architecture.
\end{abstract}

\section{Introduction}
The security of communication networks has become increasingly critical in today’s digital infrastructures, demanding robust cryptographic protection~\cite{joseph2022transitioning}. At the same time, the anticipated development of large-scale quantum computers threatens the security of standard cryptographic schemes, as quantum algorithms challenge the computational assumptions underlying widely deployed encryption protocols~\cite{shor1994algorithms, grover1996fast, jordan2018quantum}. In particular, the prospect of “harvest-now, decrypt-later” attacks motivates the development of communication systems resilient to quantum adversaries~\cite{mosca2018cybersecurity}.

Quantum key distribution (QKD) offers a complementary approach to securing communications against quantum adversaries, enabling information-theoretically secure key distribution based on the laws of quantum physics~\cite{bennett2014quantum, scarani2009security}. Over the past two decades, significant progress has been made in both theoretical security analysis and experimental demonstrations. However, large-scale deployment of QKD infrastructure requires not only secure protocols but also mature and scalable communication systems capable of operating in heterogeneous communication environments.

From an infrastructure perspective, quantum communication networks span metropolitan and inter-city links, typically covering short to medium distances, as well as space-based and other long-haul links targeting global connectivity~\cite{kimble2008quantum}. At the metropolitan scale, optical fiber is the dominant medium for QKD deployment. Fiber-based systems provide high stability and straightforward integration with existing telecom infrastructure, and several fiber-based QKD networks have already been deployed and experimentally validated in real-world conditions~\cite{peev2009secoqc, martin2024madqci, chen2021implementation}. However, fiber-based QKD commonly relies on dedicated point-to-point links, whose deployment can be costly, particularly when suitable infrastructure is not already available. In addition, fiber-based links remain inherently vulnerable to accidental cuts or deliberate disruptions of the physical network.

Free-space (FS) links offer a complementary approach to addressing some of the limitations of fiber-based QKD deployment. As FS links do not require a guided physical transmission medium, they enable flexible node placement, rapid deployment, and improved network resilience. Despite these advantages, the practical deployment of urban FS QKD links remains challenging, particularly in dense urban environments (with population densities exceeding $10^4$ inhabitants/km²). Establishing and maintaining a stable optical channel requires dedicated terminal hardware and precise alignment. Such environments also concentrate critical infrastructure, including major airports and their surrounding controlled airspace, which may impose additional constraints on optical emissions and restrict the use of high-power laser beacons for pointing and tracking. Urban propagation paths may also experience significant atmospheric turbulence and associated channel fluctuations~\cite{mostafa2012channel}.

Long-distance FS implementations may rely on higher-order adaptive optics, complex stabilization schemes, or advanced photon-detection technologies, depending on the wavelength and propagation environment~\cite{Cao2020, liao2017long, rossi2026intermodal}. While effective, such approaches increase system complexity and may limit practical deployability.

Despite these challenges, ground-based FS QKD, typically spanning tens of kilometers, has been demonstrated in several experimental settings. Cai \emph{et al.} reported a 20~km decoy-state BB84 implementation with continuous day--night operation over a rural channel~\cite{Cai2024free}, while Liao \emph{et al.} demonstrated a 53~km horizontal link over lake and non-populated areas under high-loss, satellite-like conditions~\cite{liao2017long}. Exceptional distances exceeding 100~km have also been achieved between astronomical observatories~\cite{SchmittManderbach2007, Ursin2007}. However, these long-range demonstrations were conducted in rural, low-background, or observatory environments rather than dense metropolitan corridors.

In contrast, urban FS QKD has predominantly been demonstrated over links of only a few kilometers~\cite{Hughes2002, ntanos2025practical, stathis2026practical, arteaga2025integration, picciariello2025intermodal}. More recently, longer total spans have been reported in metropolitan environments using interference-based protocols:
Cao \emph{et al.} achieved measurement-device-independent (MDI) QKD over a total span of 19.2~km, comprising two interfering links of 7.7 and 11.5~km~\cite{Cao2020}, while Li \emph{et al.} demonstrated twin-field (TF) QKD over a total span of 14.2~km, comprising two 7.1~km links~\cite{li2026free}. These demonstrations therefore correspond to the combined span of multiple shorter FS links rather than to a single extended horizontal channel. Direct urban links beyond 15~km remain scarce, with a notable recent exception being an 18~km QKD experiment at 1550~nm that relied on adaptive optics to operate across the urban channel~\cite{rossi2026intermodal}.

Table~\ref{tab:fsl_state_of_art} summarizes representative ground-to-ground FS QKD demonstrations with link distances $> 10$~km, illustrating that long direct links have predominantly been demonstrated in sparsely populated environments. {Moreover, it is evident that any ground-to-ground free-space experiment above 20 km has been conducted at altitudes much higher than sea level, where the atmosphere is considerably less dense, and thus weaker turbulence is expected.}

\begin{table}[t]
\centering
\begin{threeparttable}
\caption{Representative ground-to-ground FS QKD demonstrations with link
distances exceeding 10~km, grouped according to the reported propagation
environment.\tnote{$\dagger$}}
\label{tab:fsl_state_of_art}

\begingroup
\footnotesize
\renewcommand{\arraystretch}{0.95}
\setlength{\tabcolsep}{3pt}

\begin{tabular*}{\linewidth}
{@{\extracolsep{\fill}} l c l c c l @{}}
\toprule
Environment
& \shortstack{Distance\\(km)}
& Protocol
& \shortstack{$\lambda$\\(nm)}
& \shortstack{Altitude\\(m a.s.l.){\tnote{$\ddag$}}}
& Reference \\
\midrule

\multirow[c]{5}{*}{Urban}
& 10.2
& HD ent.-based QKD
& $\sim810$
& $\sim${200--250}
& Bulla \emph{et al.} \cite{bulla2023nonlocal} \\

& 13$^*$ {\scriptsize $(7.7+5.3)$}
& Ent.-based QKD
& $\sim702$
& $\sim${284--30}
& Peng \emph{et al.} \cite{peng2005experimental} \\

& 14.2$^*$ {\scriptsize $(7.1+7.1)$}
& TF-QKD
& $\sim1550$
& $\sim${10} 
& Li \emph{et al.} \cite{li2026free} \\

& 18
& BB84 (decoy)
& 1550
& $\sim${474--12}
& Rossi \emph{et al.} \cite{rossi2026intermodal} \\

& 19.2$^*$ {\scriptsize $(7.7+11.5)$}
& MDI-QKD
& $\sim1550$
& $\sim${10}
& Cao \emph{et al.} \cite{Cao2020} \\

\midrule

\shortstack{\textbf{Densely populated}\\
            \textbf{urban corridor}}
& \textbf{18.9}
& \textbf{BB84 (decoy)}
& \textbf{785}
& \textbf{$\sim${25--425}}
& \textbf{This work} \\

\midrule

\multirow[c]{2}{*}{Rural/lake}
& 20
& BB84 (decoy)
& $\sim780$
& $\sim${2266--2070}
& Cai \emph{et al.} \cite{Cai2024free} \\

& 53
& BB84 (decoy)
& 1550
& $\sim${3208--3242}
& Liao \emph{et al.} \cite{liao2017long} \\

\midrule

\multirow[c]{2}{*}{Observatory}
& 144
& BB84 (decoy)
& 850
& $\sim${2392--2410}
& Schmitt\add{-Manderbach} \emph{et al.} \cite{SchmittManderbach2007} \\

& 144
& Ent.-based QKD
& 710
& $\sim${2392--2410}
& Ursin \emph{et al.} \cite{Ursin2007} \\

\bottomrule
\end{tabular*}

\endgroup

\begin{tablenotes}[flushleft]
\scriptsize
\item[*] The reported distance corresponds to the total span of two
free-space links rather than a single direct channel.
\item[$\dagger$] Propagation environments are classified according to
the geographical descriptions reported in the corresponding publications.
Altitude values are given in meters above sea level (m a.s.l.); approximate
values ($\sim$) are inferred from the geographical locations of the
experimental terminals when not explicitly reported. For this work,
population densities along the crossed municipalities range from
approximately $1.5\times10^{3}$ to $1.3\times10^{4}$
inhabitants/$\mathrm{km}^{2}$.
\item[$\ddag$]{This column includes one-value entries for when both the transmitter and the receiver are at the same altitude, or two-value entries as Tx altitude -- Rx altitude.}
\end{tablenotes}

\end{threeparttable}
\end{table}

From a deployment perspective, some of these demonstrations rely on advanced detection technologies, including superconducting nanowire single-photon detectors (SNSPDs)~\cite{Cao2020, li2026free} or upconversion-based detectors~\cite{liao2017long}, as well as auxiliary optical channels for acquisition, pointing, tracking, synchronization, and adaptive-optics control~\cite{Cao2020, liao2017long, rossi2026intermodal}. The reliance on cryogenic or wavelength-conversion detection systems and additional optical channels can increase system complexity and hinder straightforward deployment, particularly at sites subject to strict optical-emission constraints.

Practical deployment imposes demanding requirements not only on the FS terminal but also on the quantum-state transmitter. Interferometric phase-encoding architectures require stable relative optical phases and are therefore sensitive to thermal and mechanical perturbations. Polarization transmitters based on multiple optical sources avoid this interferometric requirement, but differences between emitters may produce state-dependent temporal, spectral, spatial, or intensity signatures, introducing potential side channels~\cite{ko2017critical,biswas2021experimental,linale2026}. A deployment-oriented transmitter should therefore combine intrinsic stability with secure state preparation and straightforward integration into the optical terminal. This calls for a fiber-coupled architecture that derives all polarization states and decoy intensities from a single optical emitter while providing accurate intensity modulation and stable, reconfigurable polarization control. These capabilities are not always simultaneously available in  most reported field implementations.

In this work, we demonstrate a polarization-encoded decoy-state BB84 system over a single, direct 18.9~km FS link across a densely populated corridor in the Barcelona metropolitan area. The link connects two urban sites in Castelldefels and Barcelona and operates entirely within the optical-emission restriction zone of Barcelona--El Prat Airport. The system combines a stable single-emitter transmitter, room-temperature single-photon detection, low-power LED-based pointing and tracking, tip--tilt beam stabilization without higher-order adaptive optics, and independent GPS-disciplined synchronization.

The paper is organized as follows. In Sec.~\ref{sec:FSL}, we describe the FS optical and detection subsystems, emphasizing their main characteristics and their compatibility with flexible integration into quantum communication experiments. In Sec.~\ref{sec:PolTx}, we present the deployment-oriented QKD transmitter architecture and its laboratory characterization, highlighting the key design features enabling operation in this deployment scenario. Then, in Sec.~\ref{sec:Exp}, we report the system-level field results. Finally, we close the article in Sec.~\ref{Conclusions} with conclusions and discussions.

\section{Free-Space Link Infrastructure}
\label{sec:FSL}

\begin{figure}[t!]
\centering
\includegraphics[width=\linewidth]{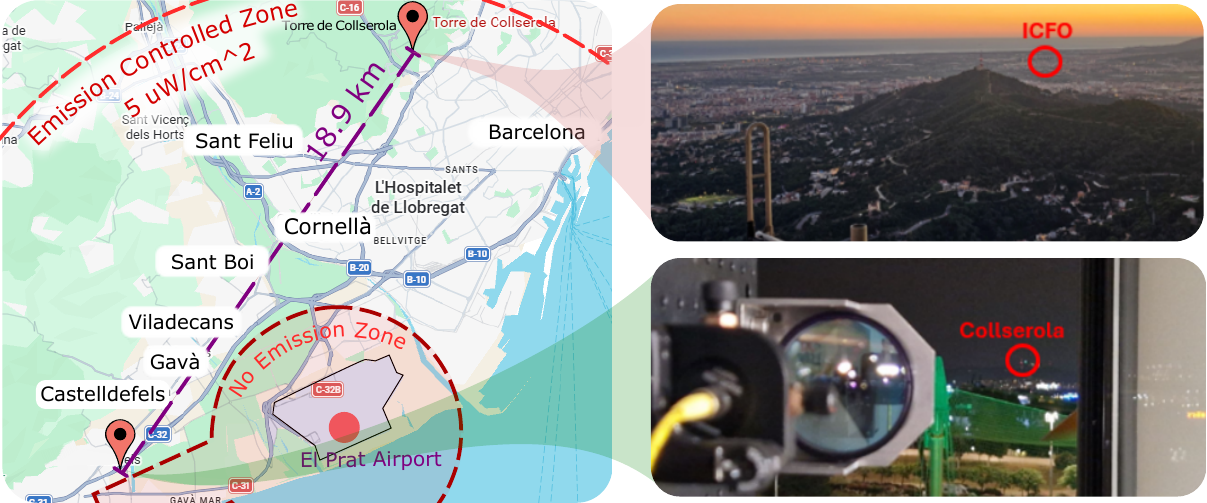}
\caption{Overview of the 18.9~km inter-city FS QKD link between the ICFO transmitter in Castelldefels (Tx-Node) and the receiver at Collserola Tower (Rx-Node). The FSL traverses the Barcelona metropolitan area, including its proximity to Barcelona--El Prat Airport. Right: views from the Rx-Node (top) and from the Tx-Node (bottom) along the line of sight. The red dashed line indicates the Laser-beam Critical Flight Zone.}
\label{fig:map}
\end{figure}

The FS link connects Castelldefels (Tx-Node) and Barcelona (Rx-Node) over 18.9~km through a continuous urban corridor crossing Gavà (\textgreater 1.5k~inhabitants/$\mathrm{km}^2$), Sant Boi de Llobregat (\textgreater 3.9k~inhabitants/$\mathrm{km}^2$), Cornellà de Llobregat (\textgreater 13.1k~inhabitants/$\mathrm{km}^2$), and Esplugues de Llobregat (\textgreater 10.5k~inhabitants/$\mathrm{km}^2$). Unlike previous long-distance demonstrations, the link connects two non-contiguous cities while traversing a densely populated urban area (Fig.~\ref{fig:map}). In addition, both nodes and the entire propagation path lie within the Laser-beam Critical Flight Zone (LCFZ) of Barcelona--El Prat Airport, where the maximum permitted irradiance is \SI{5}{\micro\watt\per\centi\meter\squared}. While this limit does not constrain the quantum signal, it restricts the auxiliary optical channels used for alignment, tracking, and stabilization, directly influencing the design of the FS infrastructure.

The Rx-Node is hosted at Collserola Tower, a major telecommunications hub in Barcelona with direct fiber connectivity to critical infrastructure. This location illustrates how inter-city FS links can be integrated into hybrid fiber--FS quantum networks through trusted nodes.

The FS infrastructure is based on a commercial research platform for FS quantum communication developed by \del{QTLabs}\add{Quantum Technology Laboratories GmbH (QTLabs)}, adapted to the installation sites, aviation-safety constraints, and operation in the 785--810~nm wavelength range. \add{The infrastructure was deployed in two stages to reduce risk. In the first stage, the telescopes, mounts, and the LED-based pointing and tracking system were installed at both sites, and the optical channel was characterized with classical light. Once it was confirmed that the link attenuation was compatible with QKD, in the second stage the tip--tilt stabilization, the PAM, and the single-photon detectors were added at the Rx-Node, and the QKD transmitter was connected to the Tx-Node terminal. Leaving aside the preparation of the laboratory spaces at both sites and the deliberate 12-month gap between the two stages, the on-site installation and alignment of the terminals took approximately 5 working days in total over the two stages, with the alignment carried out at night under good weather conditions. Because the QKD transmitter is fiber-coupled and the synchronization relies on GPS-disciplined clocks, connecting the quantum payload to the optical terminals did not require any modification of the terminal hardware.}

\subsection{Free-space link optical setup}

\begin{figure}[t!]
\centering
\includegraphics[width=\linewidth]{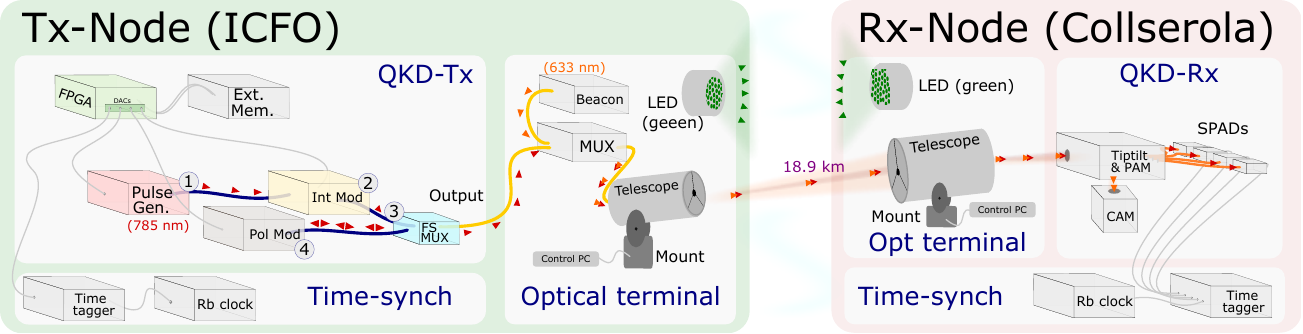}
\caption{System-level architecture of the 18.9~km FS QKD link. The transmitter node comprises a QKD transmitter (\textit{QKD-Tx}) consisting of FPGA-controlled pulse-generation (1), intensity-modulation (2), and polarization-modulation (4) units, together with a free-space multiplexer (3) that couples the prepared quantum states into a single-mode fiber. The \textit{Optical terminal} comprises the transmitting telescope, mechanical mount, and pointing, acquisition, and tracking (PAT) system, together with auxiliary beacons at 633 and 500~nm and a low-insertion-loss optical multiplexer (MUX) that combines the 633~nm beacon with the quantum signal. At the receiver node, the collected signal is stabilized and analyzed by the tip--tilt and polarization-analysis modules before detection by the SPADs. Independent GPS-disciplined rubidium clocks and time taggers provide temporal synchronization at both sites.
Yellow lines: Single Mode fiber; Blue lines: Polarization Maintaining fiber; Orange lines: Multimode fiber; Gray lines: RF electrical connections. The colored arrows indicate the different optical signal paths and their directions.
}
\label{fig:scheme}
\end{figure}

The complete system-level architecture, including the FS optical subsystem, is shown in Fig.~\ref{fig:scheme}. The optical subsystem comprises a transmitter terminal at ICFO (Tx-Node) and a receiver terminal at Collserola Tower (Rx-Node). The receiver employs a single 406~mm Ritchey--Chrétien telescope to collect both the quantum signal and the optical tracking beacon. At the transmitter, a 203~mm telescope is dedicated to tracking, while the quantum signal is launched through a fiber-coupled collimation stage based on a 75~mm achromatic lens, optimized for operation over the 780--920~nm spectral range. The collimation stage is installed adjacent to the tracking telescope to enable co-alignment of the quantum and tracking channels.

Both terminals are mounted on motorized alt-azimuth mounts, enabling automatic pointing and tracking of the FS link. To comply with the LCFZ constraints, coarse alignment and tracking are implemented using low-power, non-collimated LED beacons rather than the high-power laser beacons commonly used in FS optical communication systems. The beacon signals are detected by cameras coupled to the telescope assemblies, and a tracking algorithm continuously corrects the terminal pointing, achieving an angular accuracy on the order of one arcsecond at an update rate of approximately 1~Hz.

After propagation over the 18.9~km FS link, the transmitted beam is expected to form a 2--3~m diameter footprint at the receiver under typical atmospheric conditions. The receiver telescope collects a fraction of this light and collimates it to a beam diameter of approximately 12.5~mm. Before the beam reaches the polarization-analysis module (PAM), it is actively stabilized by a fast steering mirror (FSM) operating in a closed feedback loop with a low-power 633~nm reference beam (less than 1~\si{\micro\watt} at the Tx-Node). The reference and quantum signals propagate collinearly through the free-space channel and are jointly acted on by the FSM. Downstream of the FSM, a dichroic mirror separates the reference beam from the quantum signal. The reference beam is imaged onto a fast camera, and the measured spot displacement provides the error signal used to drive the FSM tip--tilt correction. In this way, beam wander is corrected before the quantum signal enters the polarization-analysis stage.

The stabilized quantum signal is then routed to the PAM, which provides extinction ratios exceeding $23$~dB in both measurement bases, corresponding to an ideal polarization-crosstalk contribution to the QBER below $\sim 0.5\%$. The four polarization outputs are subsequently coupled into multimode fibers and connected to commercial silicon SPAD-based single-photon detectors (with a detection efficiency of approximately 60\%)\add{.} The active tip--tilt stabilization maintains stable coupling into the detection fibers without requiring higher-order adaptive optics. This configuration enables efficient photon collection with the 406~mm aperture telescope, with an overall receiver loss below 7~dB when accounting for the polarization-analysis optics, fiber coupling \add{(which can be improved with careful alignment)}, and single-photon detector efficiency, while remaining compatible with the simplified, deployment-oriented architecture adopted in this work. 

\begin{figure}[t!]
\centering
\includegraphics[width=\linewidth]{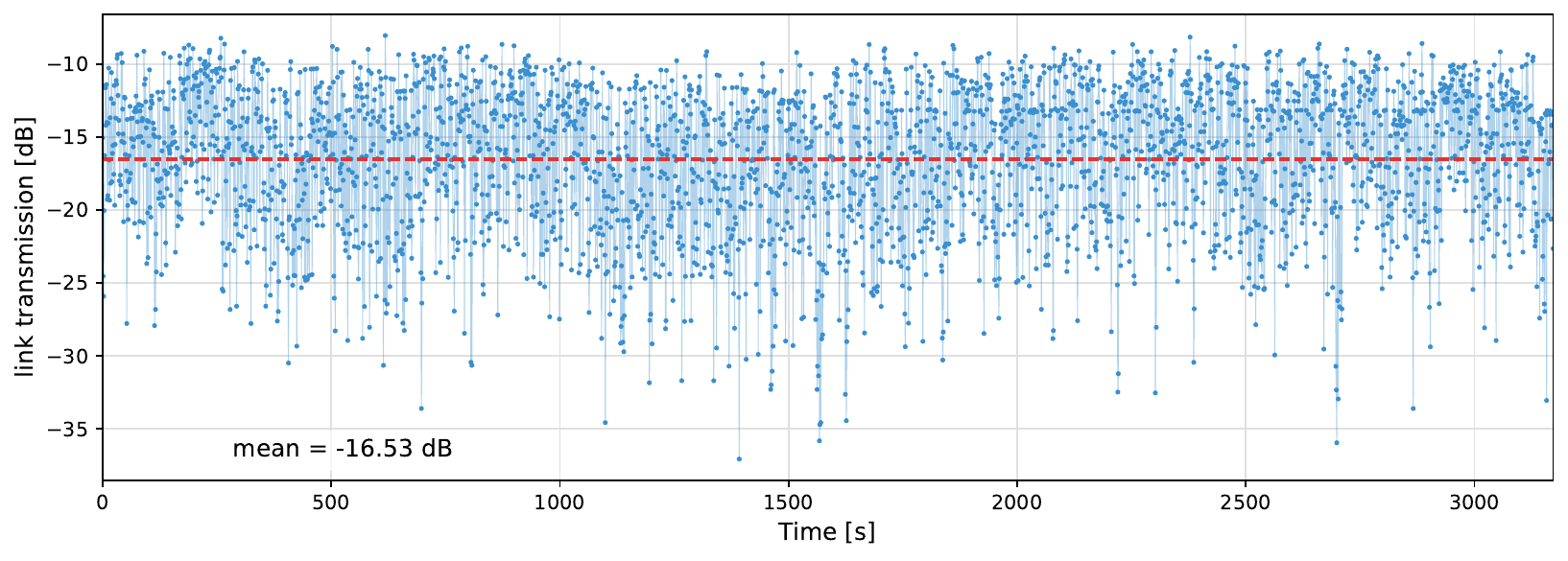}
\caption{Measurement of the quantum link transmission by transmitting a continuous-wave signal at $785$~nm from the transmitter to the receiver. The optical power was measured with a power meter at the receiver input, before the PAM. The plot shows the typical atmospheric fluctuations of the link over \del{a 5-minute interval}\add{more than 50~minutes}, illustrating the characteristic behavior of the channel during a representative measurement. The red dashed line indicates the mean transmission of \del{$-18.04$}\add{$-16.5$}~dB. \add{ This measurement was recorded during the revision of this work.}
}
\label{fig:channel_IL}
\end{figure}

Fig.~\ref{fig:channel_IL} shows a representative time evolution measurement of the FS link insertion loss. This value was obtained by transmitting a low-power continuous-wave (CW) beam at $785$~nm and monitoring every second the collected optical power over \del{a 5-minute period}\add{more than 50~minutes} at the receiver input, before the PAM. Under typical measurement conditions with favorable weather, the experimental beam experiences \add{a mean} attenuation in the range of \del{15--22~dB.}\add{16--22~dB. Besides the second-scale fluctuations caused by turbulence and residual pointing errors, this measurement shows the stability of the channel over time scales comparable to the duration of the QKD field test reported in Sec.~\ref{sec:Exp}: around a mean attenuation of 16.5~dB, the 5-min moving average varied only between 15.1 and 18.7~dB over the whole period. A dedicated long-term statistical characterization of the channel, such as those reported for classical urban free-space optical links~\cite{mostafa2012channel} and for free-space quantum channels~\cite{erven2012studying, capraro2012impact}, is beyond the scope of this work and will be the subject of future campaigns.} 

\subsection{Synchronization Architecture}

Reliable temporal synchronization between the transmitter and receiver is required for the implementation of quantum communication protocols. In our system, this is achieved using two time-tagger units, each referenced to a local rubidium clock frequency standard and synchronized to a GPS timing signal. This architecture avoids the need for an additional optical synchronization channel, reducing terminal complexity and avoiding further optical-emission constraints at the transmitter site.

Photon detection events are converted into precise time stamps referenced to a 10~MHz signal generated by the rubidium clocks. The GPS pulse-per-second (PPS) signal provides an absolute timing reference, resulting in an initial synchronization on the order of 10~ns between the two sites. During post-processing, cross-correlation of the recorded time tags with reference signals from the Tx-Node FPGA allows the relative timing to be further refined. This timing precision is sufficient to accurately identify photon detection events within the coincidence window (1 ns) used in the QKD protocol.

\section{QKD Transmitter Architecture and Laboratory Characterization}
\label{sec:PolTx}

The QKD-Tx subsystem at the Tx-Node side in Fig.~\ref{fig:scheme} illustrates the main functional blocks of the QKD transmitter for optical pulse generation, intensity modulation, and polarization modulation. The transmitter follows a deployment-oriented architecture based on the designs presented in \cite{stein2023robust, linale2026coexistence, linale2024c}, adapted in this work to operate at a wavelength of 785~nm and a repetition rate of 100~MHz. This configuration enables the use of a practical free-space QKD receiver based on high-efficiency, room-temperature silicon single-photon detectors.

First, optical pulses are generated by the direct modulation of a distributed-feedback laser (DFB) and routed through a polarization-maintaining fiber (block 1) to an intensity-modulation stage based on a Sagnac loop (block 2). This stage incorporates a phase modulator and prepares the different mean photon numbers required for decoy-state operation. The Sagnac configuration mitigates pattern-dependent effects and enables stable multilevel intensity modulation~\cite{PatterningEffectMitigation}. Polarization modulation (block 4) is implemented by controlling the relative phase between the orthogonal fast and slow axes of the PM fiber, corresponding to the horizontal, $\ket{H}$, and vertical, $\ket{V}$, polarization states. The prepared state can be expressed as $\ket{\psi} = \frac{1}{\sqrt{2}}\left(\ket{H} + e^{i\phi}\ket{V}\right)$, where the phase $\phi$ is electronically controlled to generate the required BB84 polarization states. As in the original 1550~nm implementation, the system supports polarization-encoded faint-pulse protocols and can implement both standard and efficient decoy-state BB84~\cite{Ma2005, EfficientBB84}, owing to flexible preparation of polarization and intensity states. A free-space multiplexer unit (block 3) provides a stable optical interface between the different transmitter building blocks and the system output.

Finally, system control is implemented using an RFSoC FPGA platform equipped with high-bandwidth 16-bit digital-to-analog converters (DACs) for pulse generation and modulation control. Depending on the operating mode, the transmitter can generate quantum-state and decoy-level sequences from patterns stored in the FPGA's internal memory or receive random data in real time using a commercial high-speed optical transceiver and store them in external memory. This flexible architecture supports different external entropy sources, including quantum random number generators (QRNGs) and physical true random number generators (TRNGs). Auxiliary digital outputs provide the emitted-state information required for synchronization and post-processing.

\subsection{Transmitter calibration and laboratory stability test}

Before integration with the FS communication system, the transmitter underwent laboratory characterization and stability testing. For the characterization, the transmitter was configured to generate fixed states rather than random sequences, and the drive voltages controlling the optical pulse intensity and polarization state were independently swept. The prepared quantum states were characterized using a standard fiber-based polarization-analysis setup as described in Ref.~\cite{stein2023robust}.

\begin{figure}[t!]
\centering
\includegraphics[width=\linewidth]{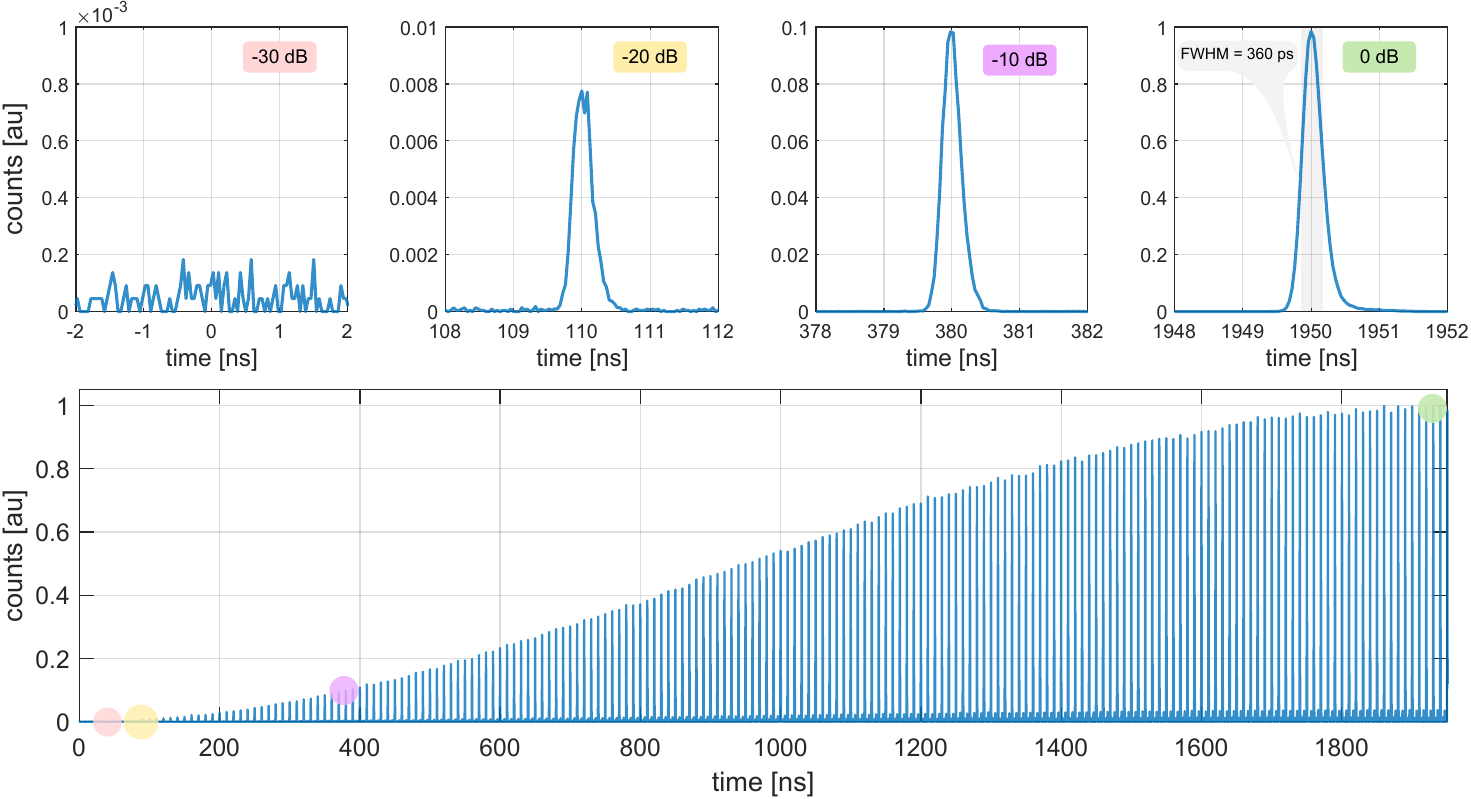}
\caption{Characterization of the intensity-modulation stage. Bottom: sequence of optical pulses with progressively varying amplitudes, obtained by selecting 200 values spanning one $V_\pi$, from maximum transmission to maximum attenuation, demonstrating analog modulation over more than 30~dB. Top: zoomed temporal profiles of representative pulses corresponding to attenuation levels on the order of (from left to right) $-30$, $-20$, $-10$, and $0$~dB. Photon counting was performed using low-jitter SNSPDs to accurately determine the pulse width.}
\label{fig:mu_sweep}
\end{figure}

The generated optical pulses exhibited a full width at half maximum (FWHM) of approximately 360~ps. By sweeping the voltage applied to the intensity-modulation stage (block 2) over 200 voltage values spanning one $V_\pi$, from maximum transmission to maximum attenuation, we verified an extinction ratio exceeding 30~dB, as shown in Fig.~\ref{fig:mu_sweep}. Based on this characterization, a fixed attenuation was introduced in the optical path, and the operating voltages were selected to produce mean photon numbers of $\mu_s = 0.45$ photons per pulse for the signal states and $\mu_d = 0.22$ photons per pulse for the decoy states, while the maximum-attenuation operating point was selected for the vacuum states.

\begin{figure}[t!]
\centering
\includegraphics[width=\linewidth]{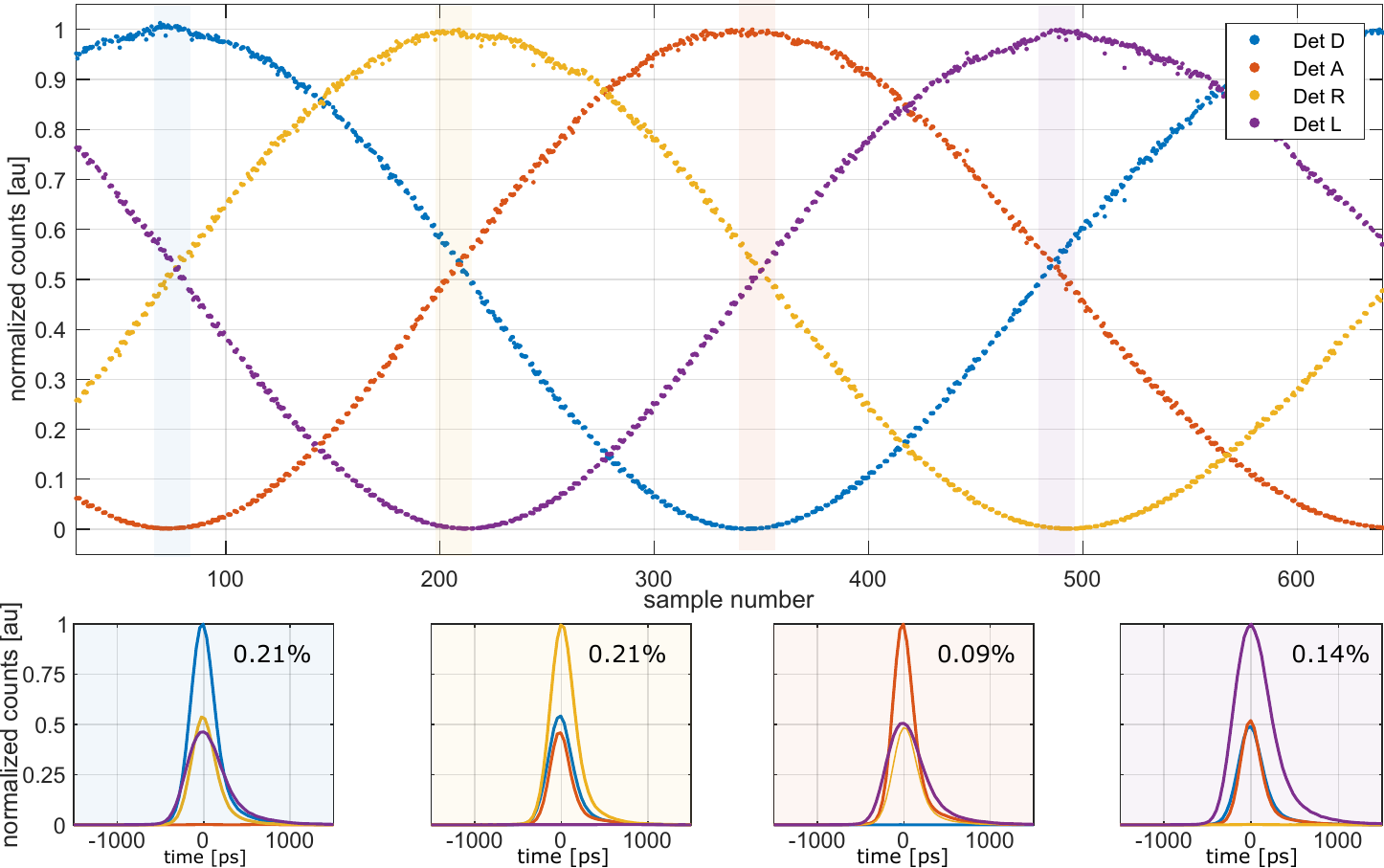}
\caption{Characterization of the polarization modulation stage. Top: detected counts projected onto the four polarization states, D, A, R, and L, as the preparation voltage is swept. The shaded regions indicate the voltage ranges selected for system calibration. Bottom: detailed view of the histograms obtained by applying the calibrated voltages to prepare the four polarization states, together with the corresponding intrinsic QBER values measured in back-to-back configuration with a fiber-based polarization-analysis setup. The mean intrinsic QBER of the transmitter is 0.16 \%.}
\label{fig:pol_sweep}
\end{figure}

Subsequently, a voltage sweep was performed on the polarization-control module (block 4) to identify the operating points required to prepare the polarization states used in the QKD protocol. The relative phase between the orthogonal polarization components of each optical pulse, $\ket{H}$ and $\ket{V}$, was controlled by the FPGA DAC, enabling precise preparation of the polarization states. The top panel of Fig.~\ref{fig:pol_sweep} shows the polarization sweep projected onto the four analysis states $\ket{D}$, $\ket{A}$, $\ket{R}$, and $\ket{L}$, corresponding to two mutually unbiased bases (MUBs). The bottom panel of the figure shows the selected operating points used for state preparation. From these projections, the intrinsic QBER associated with each prepared polarization state can be estimated, providing a lower bound for the QBER expected during full protocol operation.

The voltage sweep shown in Fig.~\ref{fig:pol_sweep} consisted of approximately 200 samples spanning one $V_{2\pi}$, with consecutive voltage points separated by approximately 60 DAC steps and each voltage level held for 3~s. The complete sweep covered only 20\% of the available DAC dynamic range. This suggests that in a future integrated implementation the system could use lower-resolution DACs, provided that the amplification gain is adjusted accordingly.

\subsection{Stability}
\label{sec:stability}
After calibration, a transmitter stability test was performed under operating conditions representative of the QKD protocol. This test was intended to verify whether a single system calibration remained valid over the duration of a complete FS measurement session.

The transmitter was first configured to generate the full set of quantum states required for a decoy-state BB84 protocol with two decoy levels, using the operating points obtained from the fixed-state calibration. The stability test consisted of a continuous 15-hour back-to-back measurement. 
During this period, the transmitter randomly generated polarization and intensity states according to a standard decoy-state BB84 protocol. The four polarization states were prepared with equal probability (25\% each), while the signal, decoy, and vacuum intensity levels were selected with probabilities of 50\%, 25\%, and 25\%, respectively. Note that this hardware implementation enables flexible choice of the polarization and intensity probabilities.

\begin{figure}[t!]
\centering
\includegraphics[width=\linewidth]{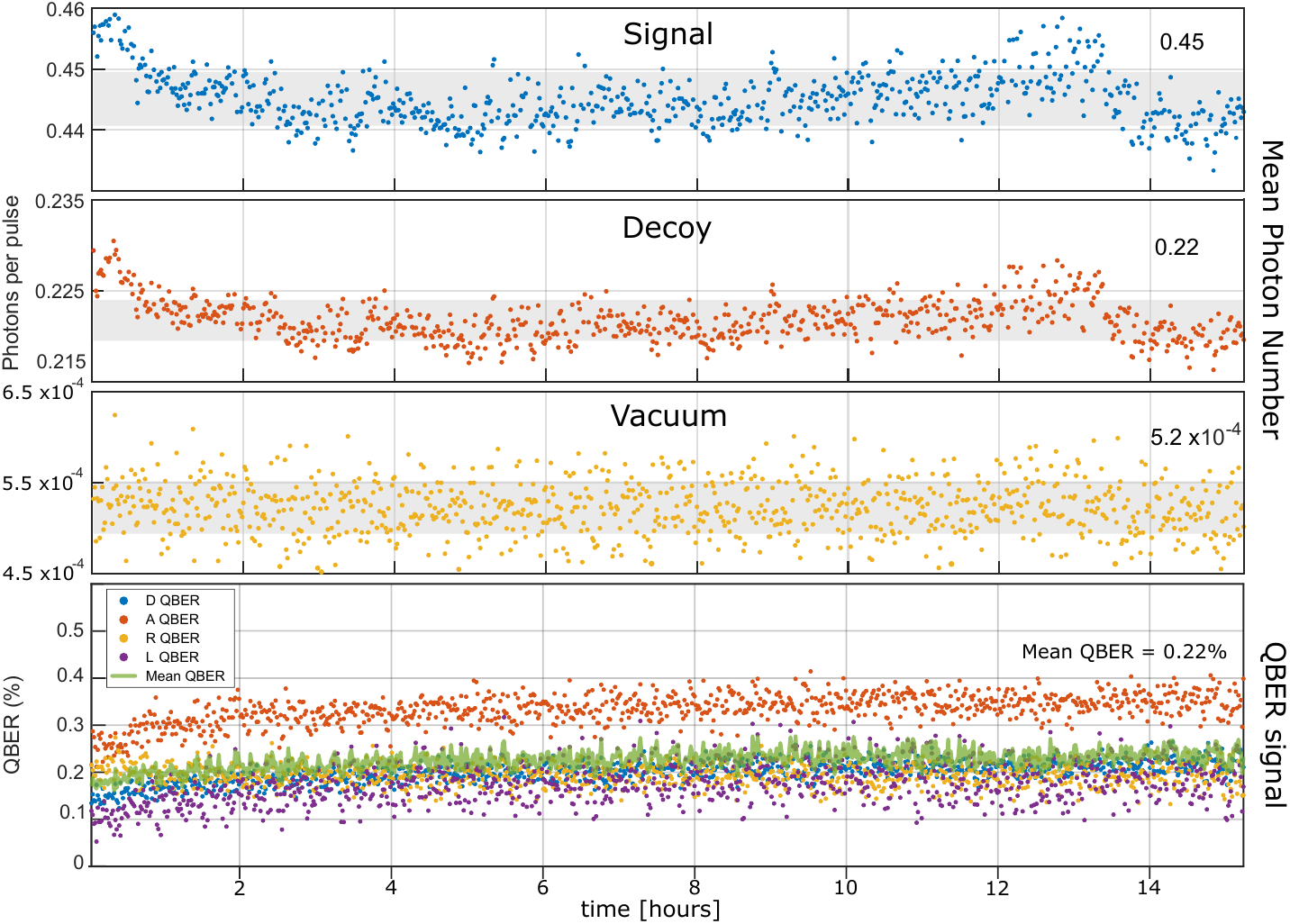}
\caption{Long-term stability of the transmitter under laboratory conditions in a back-to-back configuration during 15~hours of continuous operation. The transmitter generated randomly selected polarization and intensity states according to a standard decoy-state BB84 protocol. Top: evolution of the mean photon number for the signal, decoy, and vacuum states, with average values of 0.445, 0.222, and $5.23\times10^{-4}$ photons per pulse, respectively. The grey zone represents one-standard deviation region around signal, decoy, and vacuum mean values. Bottom: QBER for the four polarization signal states, with an average value of $(0.22 \pm 0.02)\%$. Each point corresponds to a 100~ms acquisition recorded once per minute.
}
\label{fig:stability}
\end{figure}

To monitor the stability of the intensity and polarization states over time, 100~ms acquisition windows were recorded once per minute, corresponding to approximately $10^{7}$ emitted states per acquisition. Results of the temporal evolution of the mean photon number for each intensity class are shown in the top panels of Fig.~\ref{fig:stability}. The mean photon numbers remained stable throughout the duration of the experiment, with only minor fluctuations around their nominal values, demonstrating that the intensity-modulation stage can maintain its operating point over extended periods without active feedback control.

The corresponding evolution of the intrinsic QBER over the same 15-hour period is shown in the bottom panel of Fig.~\ref{fig:stability}. The measured average QBER under protocol-representative operating conditions was $(0.22 \pm 0.02)\%$, demonstrating stable polarization encoding and low intrinsic error over extended continuous operation.

\section{Experimental results}
\label{sec:Exp}
\subsection{Fixed-Pattern Transmission and QBER Estimation}

As an initial system-level test, a fixed sequence of the $\ket{D}$, $\ket{A}$, $\ket{R}$, and $\ket{L}$ polarization states was transmitted with a mean photon number of 1.5 photons per pulse, approximately 2--3 times higher than the value typically used for QKD operation. The increased signal level reduced the relative contribution of background and detector dark counts, allowing the effective QBER associated mainly with state-preparation imperfections, receiver polarization crosstalk, and transmitter--receiver misalignment to be evaluated. Based on the independent laboratory characterizations, the expected QBER contributions from the transmitter and the PAM are approximately 0.2\% and below 0.5\%, respectively. We note that the polarization-state set is calibrated independently at the transmitter, while Tx--Rx polarization alignment is performed by matching one state from each basis to the corresponding PAM outputs. In this experiment, the alignment was performed manually, although the same procedure can be readily automated, as demonstrated in Ref.~\cite{linale2026coexistence}.

\begin{figure}[t!]
\centering
\includegraphics[width=\linewidth]{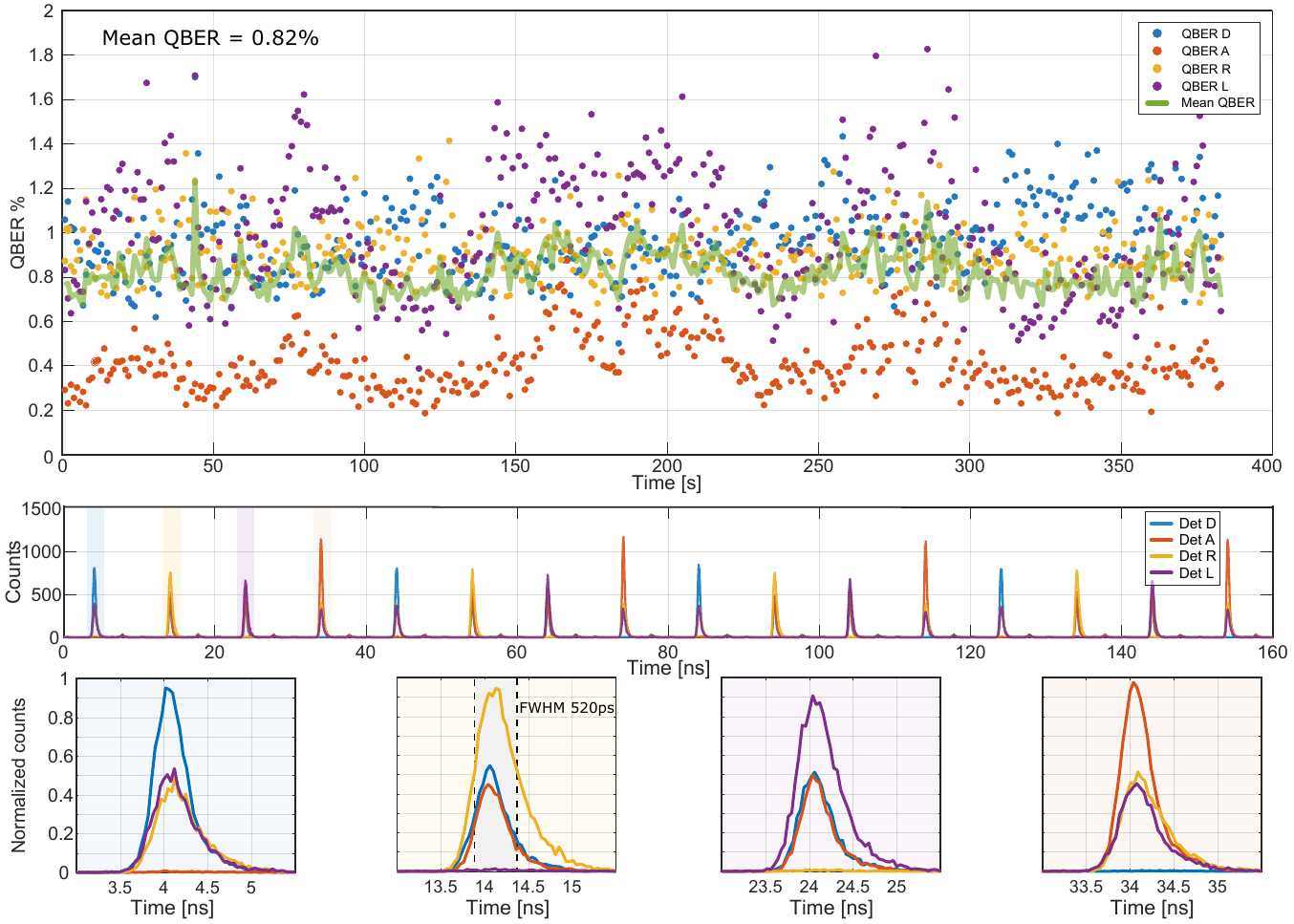}
\caption{System-level performance. Top: Measured QBER for the fixed polarization sequence (D, A, R, L), over 6 minutes at a mean photon number of $1.5$ photons per pulse. The green line shows the mean QBER, with an average value of 0.82\% over the measurement, whereas colored dots show individual QBER measurements with mean values of D: 0.99\%, A: 0.40\%, R: 0.91\%, and L: 1.00\%. Middle: Selected one-second count histogram at Rx-Node corresponding to the first second of QBER measurement for the four polarization states. Differences in peak height are primarily attributed to variations in the free-space-to-fiber coupling efficiency at the PAM module. Bottom: Enlarged temporal profiles corresponding to the shaded regions in the middle panel, showing the normalized count distributions of the four polarization states.
}
\label{fig:QBERHVDA}
\end{figure}

The fixed transmission pattern also enables reconstruction of the individual polarization-state histograms, providing information on their temporal distribution and timing jitter after synchronization. Fig.~\ref{fig:QBERHVDA} shows the QBER over a $380$~s measurement interval together with a detection histogram integrated over $1$~s. The measured mean QBER was $0.82\%$. The reconstructed optical pulses had a temporal width of approximately $500$~ps FWHM; therefore, we set a $1$~ns valid-count digital gating window around each histogram peak. The measured QBER values for the individual polarization states remained stable throughout the measurement interval.

\subsection{QKD field test and parameter estimation}
Next, we emulated the QKD protocol by repeatedly transmitting a 256-symbol pseudorandom sequence. This measurement provided the parameters required to estimate the secret key rate (SKR) for a standard decoy-state BB84 protocol with two decoy levels. The polarization-state and intensity-selection probabilities were set to the same values used in the laboratory stability characterization described in Sec.~\ref{sec:stability}. To evaluate the stability of the complete system under field conditions, the sequence was transmitted continuously for 40~min. Before the measurement session, the FS link attenuation was measured over a 5~min interval, yielding an average value of 18.5~dB.

\begin{figure}[t!]
\centering
\includegraphics[width=\linewidth]{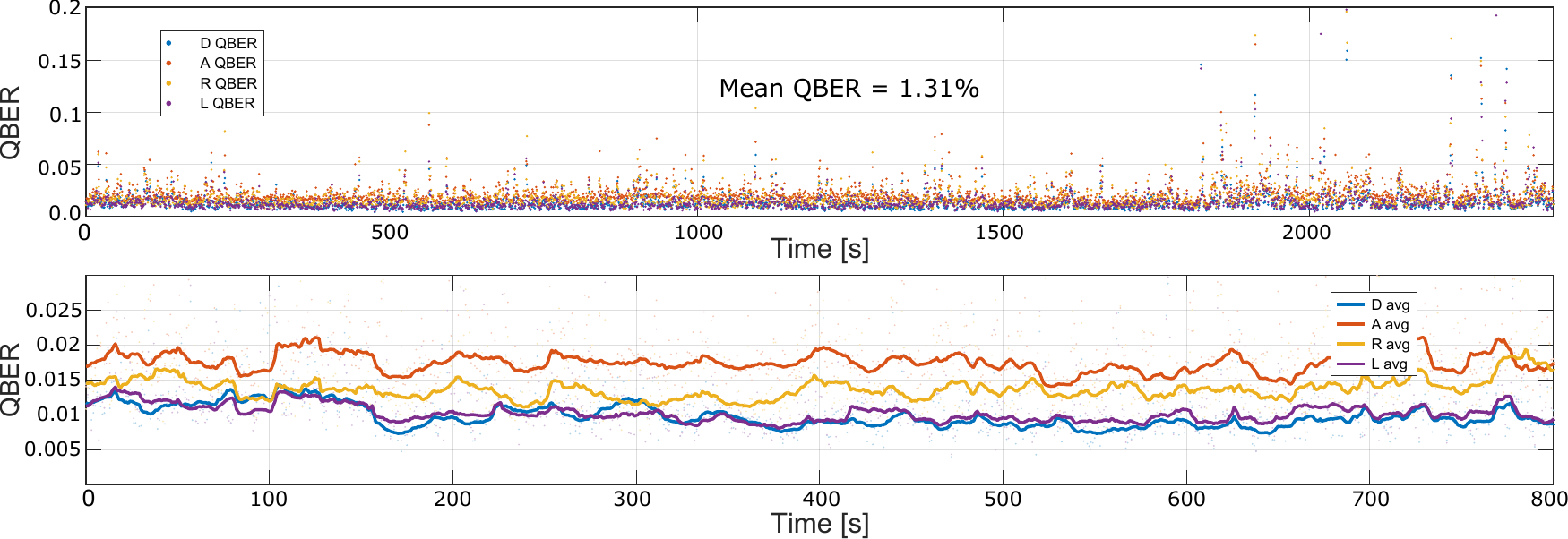}
\caption{Signal-state QBER measured continuously with 1~s resolution during the 40~min FS link test. Top: direct QBER estimates for each polarization channel over the full measurement interval. Bottom: detailed view of the first 800~s. The solid lines indicate the 20~s detection-weighted moving average QBER per polarization state. The weighted QBER remains stable despite the observed fluctuations in the direct 1~s estimates.}
\label{fig:FS_QKD_test}
\end{figure}

During post-processing, the QBER was evaluated for each polarization state, together with the detection gains (i.e., the probability of detection at the receiver per emitted pulse) for the signal, decoy, and vacuum intensity levels. Fig.~\ref{fig:FS_QKD_test} shows the temporal evolution of the signal-state QBER for each polarization channel with 1~s resolution. Occasional increases in QBER were observed and were associated with intervals of reduced channel transmission and, consequently, a lower signal-to-background count ratio. Since the number of detected events varied with the channel transmission, the mean QBER was calculated as a detection-weighted average rather than as the arithmetic mean over the individual acquisition intervals:

\begin{equation*}
\mathrm{QBER}_{i}
=
\frac{\displaystyle\sum\limits_{t} \mathrm{QBER}_{i}(t)\,
\Bigl(N_{i}(t)+N_{i^\perp}(t)\Bigr)}
{\displaystyle\sum\limits_{t}
\Bigl(N_{i}(t)+N_{i^\perp}(t)\Bigr)}
\end{equation*}
where $N_i(t)$ and $N_{i^\perp}(t)$ are the numbers of detection events recorded for state $i$ and its orthogonal state $i^\perp$, respectively, during the $t$-th 1~s acquisition interval. Thus, the resulting QBERs were $0.99\%,1.79\%,1.35\%$, and $1.07\%$ for the states D, A, R, and L, respectively, with an overall weighted mean QBER of $1.31\%$. This value is representative of the QBER expected after basis sifting in a complete QKD implementation. The bottom panel shows the first 800~s of the measurement and indicates that the 20~s detection-weighted moving-average QBER remains stable for all polarization states. \add{Since the information is encoded in polarization, residual pointing errors do not by themselves rotate the received polarization states. Their effect on the QBER is indirect: beam wander that is not fully compensated by the tip--tilt loop reduces the coupling into the detection fibers~\cite{dikmelik2005fiber} and therefore lowers the signal-to-background ratio, in the same way as atmospheric attenuation does.}

\begin{figure}[t!]
\centering
\includegraphics[width=\linewidth]{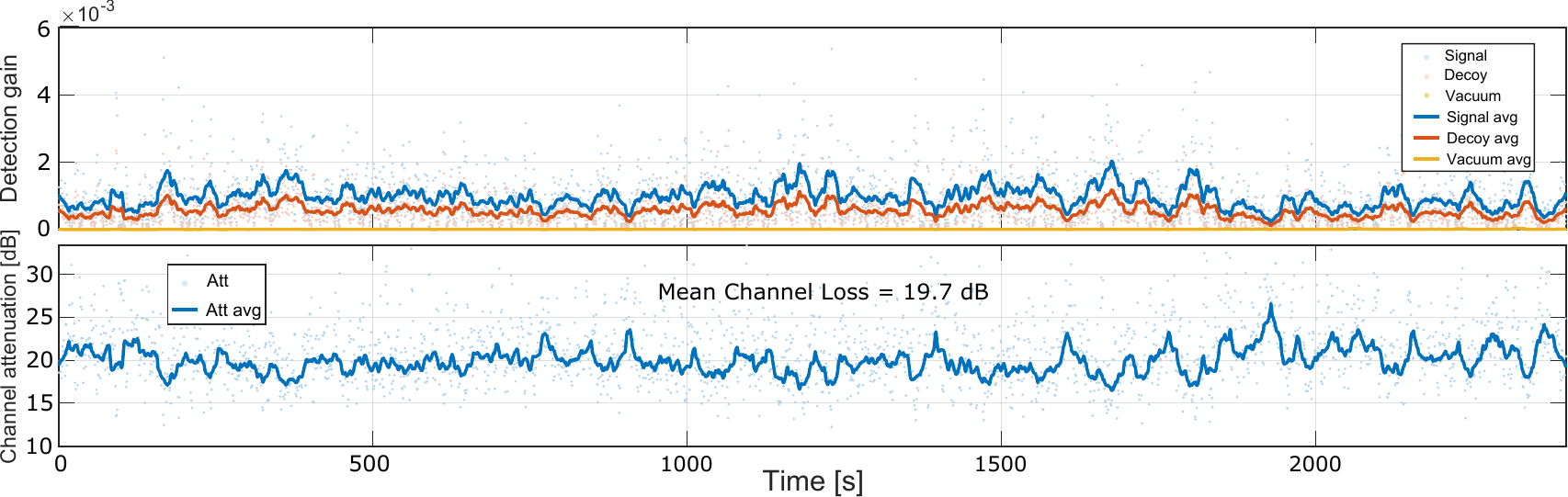}
\caption{Characterization of the FS link during the 40~min QKD emulation test. Top: detection gains for the signal, decoy, and vacuum states evaluated over 1~s integration intervals; solid lines show the corresponding 20-point moving averages. Bottom: channel attenuation estimated from the detected signal-state counts after accounting for the PAM insertion loss and detector efficiency; the solid line indicates the 20-point moving average.}
\label{fig:FS_gain}
\end{figure}

As part of the characterization, the detection gains of the signal, decoy, and vacuum states used for the secure key rate estimation~\cite{lim2014concise} were evaluated over the same 1~s integration intervals. Fig.~\ref{fig:FS_gain} shows the temporal evolution of these gains together with the link attenuation inferred from the measured signal-state detection values.

The overall gains were $Q_{\mathrm{signal}} = 0.97\times10^{-3}$, $Q_{\mathrm{decoy}} = 5.67\times10^{-4}$, and $Q_{\mathrm{vacuum}} = 1.40\times10^{-5}$. The channel attenuation was inferred from the detected signal-state photon counts after accounting for the insertion loss of the PAM and the detector efficiency. \add{Over the 40-min session, the inferred attenuation had a mean value of 19.7~dB. Although the channel attenuation cannot be measured directly during the QKD transmission with the present setup, this mean value is consistent with the one measured with the CW beam immediately before the session (18.5~dB over a 5-min measurement).}

\begin{figure}[t!]
\centering
\includegraphics[width=\linewidth]{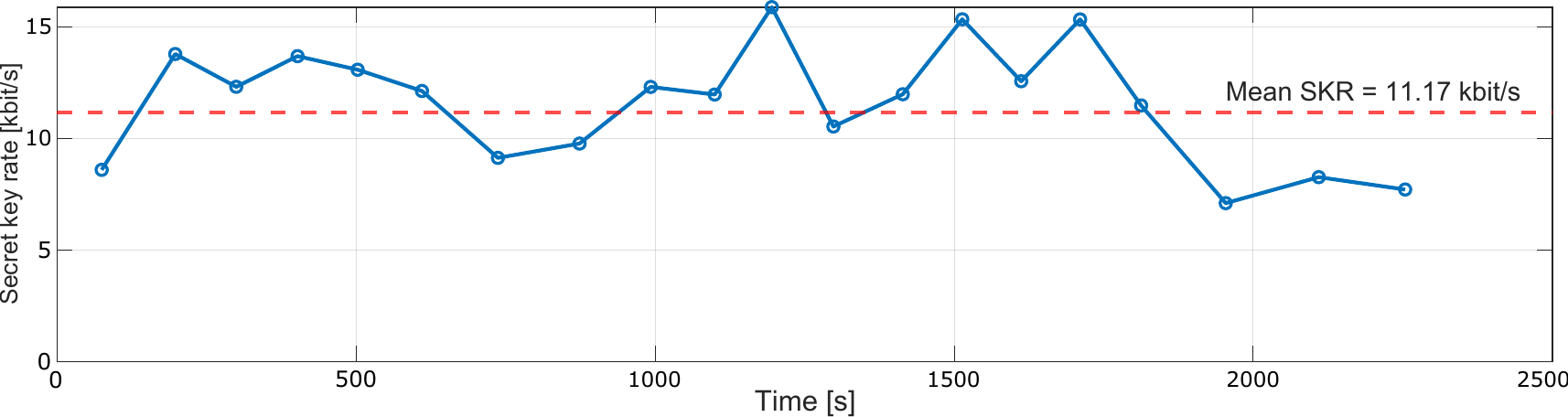}
\caption{Finite-size secure key rate evaluated throughout the measurement using target-based post-processing blocks. Each block is closed once $n_{\mathrm{key}}=2\times10^{6}$ sifted detections are accumulated in the diagonal ($D/A$) key basis. The average acquisition time per block was $116.4~\mathrm{s}$, yielding an average finite-size secure key rate of $11.17~\mathrm{kbit\,s^{-1}}$. The dashed line indicates the mean SKR.}

\label{fig:FS_SKR}
\end{figure}

Finally, we evaluate the finite-size secret key rate following the analysis of Lim \emph{et al.}~\cite{lim2014concise}. The post-processing blocks are defined by a target of $n_{\mathrm{key}}=2\times10^{6}$ sifted detections in the diagonal ($D/A$) basis, which is used for key generation, while the circular ($R/L$) basis is used for parameter estimation. The four BB84 polarization states are prepared with equal probability, $p_D=p_A=p_R=p_L=0.25$, corresponding to basis-selection probabilities $p_{DA}=p_{RL}=0.5$. The signal, weak-decoy, and vacuum intensities are $\mu=0.45$, $\nu=0.22$, and 0, respectively, with probabilities $p_{\mathrm{signal}}=0.5$, $p_{\mathrm{decoy}}=0.25$, and $p_{\mathrm{vacuum}}=0.25$. Using $\epsilon_{\mathrm{sec}}=10^{-10}$, $\epsilon_{\mathrm{cor}}=10^{-15}$, and an error-correction efficiency of $f_{\mathrm{EC}}=1.16$, we obtain an average finite-size SKR of 11.17~kbit/s. The temporal evolution of the SKR over the 40-minute measurement is shown in Fig.~\ref{fig:FS_SKR}.

\section{Conclusion and discussion}
\label{Conclusions}

We have demonstrated the consistent operation of an inter-city FS QKD link over a densely populated area under realistic field conditions using deployment-oriented hardware. The system maintained a mean QBER of $1.31\%$, while the measured channel parameters supported a finite-size secret key rate above 11~kbit/s. Beyond the achieved performance, the system architecture was designed to facilitate practical integration and deployment. The fiber-coupled transmitter, independent synchronization architecture, low-power pointing and stabilization channels, and room-temperature detection system allow the quantum payload to be interfaced with the free-space optical terminal without imposing stringent requirements on its hardware or operation. This modular approach provides a practical route toward hybrid fiber--FS quantum networks, in which FS links can complement existing fiber infrastructure and provide connectivity between geographically separated network nodes. In particular, the use of low-power auxiliary optical channels makes the architecture well suited for operation in environments subject to strict optical-emission constraints, such as densely populated areas in the vicinity of airports.

Beyond the specific QKD implementation demonstrated here, the measured stability, polarization-transmission fidelity, and channel loss of the deployed infrastructure are compatible with the implementation of other quantum-communication protocols and experiments. The demonstrated urban link therefore provides a platform for evaluating quantum technologies under realistic propagation conditions and will support further FS quantum-communication experiments.

The infrastructure was operated over a three-month period across multiple measurement campaigns. Comparable system performance was obtained on different days despite variations in atmospheric conditions and channel attenuation. Across these campaigns, the measured attenuation ranged from 16 to 22~dB, while the corresponding QBER remained between $1\%$ and $3 \%$. These observations indicate that the reported performance is reproducible across independent measurement sessions and under varying propagation conditions. \add{A full characterization of the long-term behavior of the link is still required, including attenuation statistics over months and seasons, the correlation between residual pointing errors and QBER, and continuous unattended QKD operation over days, including daytime. Channel measurements of this kind have been reported for classical urban free-space optical links~\cite{mostafa2012channel}, and continuous free-space QKD operation has been demonstrated over day--night cycles~\cite{Cai2024free} and over several days on a short urban link~\cite{peloso2009daylight}. Extending these studies to a long inter-city urban channel is the natural next step for the infrastructure presented here.}

\section*{Author contributions\del{ and Disclosures}}

I.L.G. and N.L. developed the QKD transmitter, led the measurement campaigns carried out with the system, performed the post-processing and analysis of the acquired data, and wrote the manuscript. A.E.B. implemented the FPGA adaptations required for operation at the selected wavelength. I. K-A. performed secure key rate simulations under different operating conditions to support the selection of the system parameters. M.C.iC. contributed to the integration of the QKD transmitter and its deployment within the free-space link infrastructure, and performed QKD simulations to support the field trials. E.A.O. and J.F. carried out the calibration and alignment of the receiver and tip--tilt stabilization system throughout the measurement campaigns. G.P., A.S.B, A.G., and I.K-A. performed QKD simulations and provided experimental support at both the transmitter and receiver stations during the field trial. V.P. supervised the work. All authors contributed to the installation, preliminary testing, and calibration of the FS research platform deployed between Castelldefels and Barcelona. All authors reviewed the manuscript and contributed corrections and revisions.
\del{I.L.G. and V.P. are listed as co-inventors on a patent application related to the content of this paper.}

\section*{\add{Disclosures}}
\add{I.L.G. and V.P. are listed as co-inventors on a patent application related to the content of this paper. The remaining authors declare no conflicts of interest.}

\section*{\add{Data availability}}
\add{Data underlying the results presented in this paper are not publicly available at this time but may be obtained from the authors upon reasonable request.}

\section*{Funding}
EU Horizon Europe Research and Innovation program (EuroQCI Spain - Project no. 101091638); Generalitat de Catalunya; \del{European Union (QCORE, 101145906)} \add{European Commission Horizon Europe: Marie Sklodowska-Curie Actions, QCORE, 101145906}. Views and opinions expressed are however those of the author(s) only and do not necessarily reflect those of the European Union or European Research Executive Agency. Neither the European Union nor the granting authority can be held responsible for them. A.S.B. acknowledges support by the predoctoral program FI-STEP (2025 STEP 00008), which is backed by the Secretariat of Universities and Research of the Department of Research and Universities of the Generalitat of Catalonia, as well as the European Social Plus Fund. G.P. acknowledges funding from the \del{Euro pean}\add{European} Union’s Horizon Europe research and innovation program under the Marie Skłodowska-Curie Grant Agreement No. 101081441. A.V. acknowledges financial support from the Ram\'on y Cajal Fellowship RYC2023-043066-I, and Grant No. \del{PID2024:160393NB-I00}\add{PID2024-160393NB-I00}.

\bibliography{refs_resub}

\end{document}